\documentclass{article} 
\usepackage[frozencache,cachedir=minted-cache]{minted}
\usepackage{minted}
\usepackage{microtype}
\usepackage{graphicx}
\usepackage{subcaption}
\usepackage{booktabs} 

\usepackage{hyperref}
\usepackage{xurl}

\usepackage[accepted]{icml2026}

\usepackage{amsmath}
\usepackage{amssymb}
\usepackage{mathtools}
\usepackage{amsthm}

\usepackage[capitalize,noabbrev]{cleveref}

\usepackage[textsize=tiny]{todonotes}

\icmltitlerunning{GPU Fingerprinting for Location Verification}

\begin{document}

\twocolumn[
  \icmltitle{GPU Fingerprinting for Location Verification}



  \icmlsetsymbol{equal}{*}

  \begin{icmlauthorlist}
    \icmlauthor{Wayne Tee}{pivotal}
    \icmlauthor{Jonathan Happel}{tampersec}
  \end{icmlauthorlist}

  \icmlaffiliation{pivotal}{Pivotal Research}
  \icmlaffiliation{tampersec}{TamperSec}

  \icmlcorrespondingauthor{Wayne Tee}{waynetee37@gmail.com}

  \icmlkeywords{Compute Governance, GPU Fingerprinting, Location Verification, Physically Unclonable Functions}

  \vskip 0.3in
]



\printAffiliationsAndNotice{}  

\begin{abstract}
  Robust governance of GPU chips is important for mitigating risks from unauthorized development of advanced AI models. Current methods for monitoring chip location rely on ping-based protocols backed by cryptographic keys stored on-chip. However, these keys can potentially be extracted by adversaries with physical access, compromising the location verification protocol. We address this vulnerability by proposing the use of hardware fingerprints rather than keys to identify GPUs during location verification. In addition, we develop a proof-of-concept GPU fingerprinting methodology that achieves up to 100\% re-identification accuracy in small-scale tests.
\end{abstract}

\section{Introduction}
To prevent the unauthorized use and development of advanced AI models, it is essential to have strong governance of the specialized GPU chips required to train and run them \cite{baker}. Current methods of monitoring chip location include the use of a delay-based protocol \cite{Brass_Aarne_2024}. In this method, a signal is sent from our server to the chip and the response time is measured. By considering the time interval and the speed of light, we can calculate the maximum distance between the chip and our server. By using multiple, globally distributed servers, the location of the GPU can be triangulated.

The security of this method requires the ability to verify that the response is indeed coming from the intended GPU \cite{Brass_Aarne_2024}. This prevents an adversary from relocating the actual chip while leaving a decoy in place to respond to the challenges. Current methods of identification rely on attestation provided by cryptographic keys stored on-chip as part of Nvidia's Confidential Compute stack. However, Confidential Compute is not designed to be resistant against advanced attacks by adversaries with physical access \citep[pp.~14--15]{nvidia}. The keys could hence be potentially extracted using tools like Focused Ion Beams \cite{gousselot} and Laser Scanning Microscopes \cite{krachenfels}, compromising the security of the protocol.

We therefore make two contributions. First, we show how location verification can be performed using fingerprint-based device identification instead of relying on keys. Secondly, we propose and evaluate one such fingerprinting function as a proof-of-concept.

\section{Fingerprint-Based Device Identification}
Prior work on GPU fingerprinting methods and Physically Unclonable Functions (PUFs) has shown that GPUs are not perfectly identical \cite{forlin,hohentanner,Laor_2022,Laor_Oren_2025}. Rather, the manufacturing process produces hardware differences between chips that can be measured by fingerprinting functions in order to identify and authenticate chips.

We propose using these functions to secure the location verification process as follows: Before chips are sold, they go through a registration phase where the fingerprinting functions are run on each chip and the results recorded. Post-sale, during the verification phase, trusted servers will monitor these chips by periodically requesting for the fingerprinting functions to be rerun. By ensuring that the resulting fingerprints match and the response time is within the expected range, the identity of the chips and their locations can be verified.
\vskip 0.2in

\begin{figure}[h]
    \centerline{\includegraphics[width=\columnwidth]{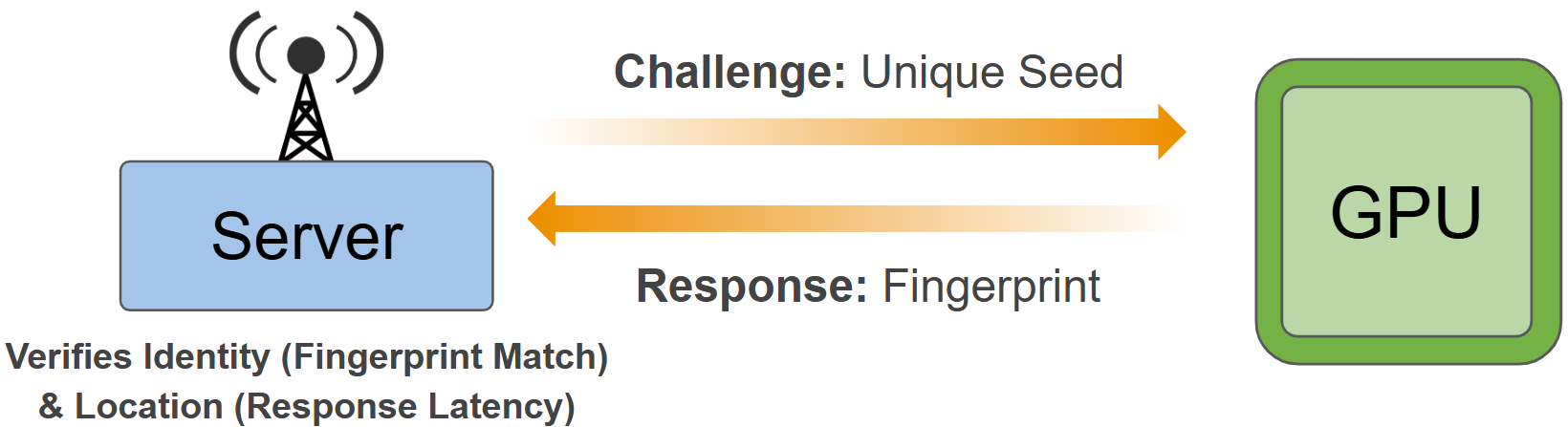}}
    \vskip 0.4in
    \caption{
      To verify the location of a GPU, the server sends a unique challenge to the chip which responds with the corresponding fingerprint. The server then authenticates the chip via the fingerprint and determines its location via the response latency.
    }
    \label{setup}
\end{figure}

\subsection{Requirements}
There are several considerations in the choice of a fingerprinting function. Firstly, the function must be both consistent and accurate. Repeated measurements on the same GPU must be consistent, while measurements across GPUs must show enough variation to distinguish the chips. Secondly, the function must be easily deployable, without the need to alter the chip manufacturing process to inject variations or the need for additional measurement devices. Software-based fingerprinting functions have an advantage here as they can be deployed immediately.

If only a single fingerprint is recorded per GPU, an adversary would be able to store the result after the first challenge, and then relocate the chip while resending the same stored result in subsequent challenges. We hence need a parameterized fingerprinting function that accepts an arbitrary seed as input and returns a fingerprint unique to the particular seed and GPU it was run on.

In this parameterized setup, the registration phase first generates a set of random numbers to use as seeds. For each GPU, the fingerprinting function is run across seeds to obtain multiple fingerprints. During the verification phase, a unique seed is selected by our server to be sent to the chip and the resulting fingerprint is compared. The uniqueness of this seed ensures that an adversary cannot simply resend past responses. Seeds must also be unpredictable to ensure that an adversary cannot guess the seeds beforehand and pregenerate responses. This minimally requires that the number of potential seeds is large enough to prevent an adversary from enumerating them all.

Unique to this context, the time required for the fingerprinting function is important. The expected response time is the sum of the round-trip signal propagation time and fingerprint computation time. If the runtime of the fingerprinting function is inconsistent, an adversary can move the chip while obscuring the increased latency within the noise in the timing of the fingerprinting process. Even if the duration of the fingerprinting function is consistent, an adversary might be able to optimize the function or use techniques such as overclocking to obtain the result more quickly. This added time margin would allow them to relocate the chip while maintaining the same overall response time.

Finally, the overall security of the system requires that an adversary cannot produce the fingerprint within the expected response time without live access to the chip. This is a difficult condition, requiring that an adversary cannot simulate, predict, or otherwise model the chip sufficiently well to compute the fingerprint, even if they have access to the fingerprinting function and the responses of other GPUs on the same seed.

\section{Proof-of-Concept Fingerprinting Function}
In this section, we present a proof-of-concept fingerprinting function. Hohentanner et al. \yrcite{hohentanner} previously showed that GPUs can be fingerprinted using atomic operations. In their \textit{atomicIncrement} approach, multiple threads race in parallel to read and increment a global counter. The values read by each thread reflect the order in which they manage to access the counter. As this order varies across GPUs, it can be saved and used as a fingerprint.
\vskip 0.1in
\begin{figure}[h]
\begin{minted}[fontsize=\fontsize{8}{10}, frame=lines,fontseries=b]{cpp}

int globalCounter = 0;

void fingerprint(int seed){
  warmup();
  for(int r = 0; r < N_ROUNDS; r++){
    if(currentThread == getRandomThread(seed)){
      randomDelay(seed);
      // Race condition
      int v = atomicAdd(globalCounter, 1);
      results[r] = v;
    }
    if(r % SYNC_INTERVAL == 0){
      synchronizeGPU();
    }
  }
}

\end{minted}
    \caption{
      Our improved fingerprinting function (simplified). This function is run in parallel across the streaming multiprocessors (SMs) that make up the GPU, producing a unique fingerprint.
    }
    \label{code}
\end{figure}

We make several improvements to this approach in order to maximize the signal-to-noise ratio of the resulting fingerprints. The refined algorithm is shown above. It includes a warm-up function and repeats the measurement across multiple rounds. During execution, the function is run in parallel on each of the streaming multiprocessors (SMs) that make up the GPU. We ensure that within each SM, only a single thread participates in the race at a time. The GPU is also periodically synchronized to perturb the results further. Results from all the SMs are concatenated together to form the final fingerprint.

We also convert the algorithm into a parameterized function. This is done by accepting a seed as input and using it to initialize a pseudorandom number generator. This is then used to select the participating thread within each SM as well as select a random delay that is then fed into a custom delay function. Note that both the delay and the thread selection vary across rounds and across SMs, maximizing the amount of signal that we obtain.

The code shown above is a simplified version that omits the full implementation details, which are essential to achieving the required accuracy. The full code and data can be found at \href{https://github.com/waynetee/gpu-fingerprinting}{https://github.com/waynetee/gpu-fingerprinting} for interested readers to refer to. Due to the proprietary nature of GPUs, the function was developed through rounds of iterative improvement guided by empirical testing rather than a deep mechanistic understanding of chip internals.

\section{Evaluation}
Using the vast.ai cloud provider, we evaluated the function across 24 Nvidia H200 GPUs, 2 seeds, and 10 runs each to obtain 480 runs in total. Each run takes about 2.9 seconds and a sample of the results is shown below:

\begin{figure}[h]
    \centerline{\includegraphics[width=\columnwidth]{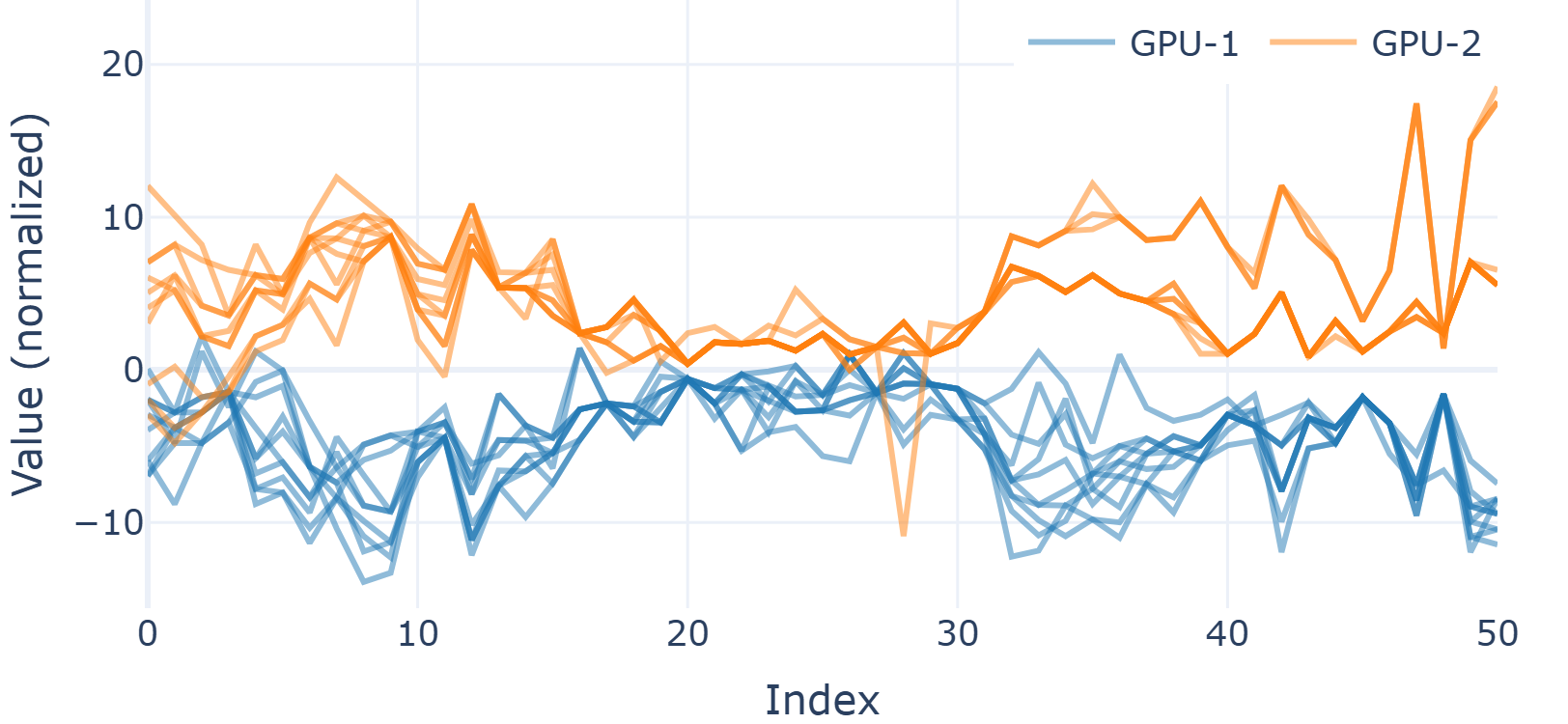}}
    \caption{
      Segment of fingerprints collected over 2 GPUs, 10 runs each with the same seed. Values have been normalized by subtracting the element-wise mean.
    }
    \label{fingerprints}
\end{figure}

\subsection{Within vs. Cross-GPU Fingerprint Distances}
The fingerprints above show clear separation between the two GPUs. However, even within the same GPU, fingerprints collected across different runs do show some amount of noise. To ensure that we can accurately identify GPUs, this within-GPU variation must be lower than the cross-GPU separation. We quantify this by first defining the distance between 2 fingerprints as the sum of absolute differences over each element (L1 distance). We then plot the distances of all pairs of fingerprints within GPUs and across GPUs on the histogram below, which shows results aggregated across both seeds.

\vskip 0.1in
\begin{figure}[h]
    \centerline{\includegraphics[width=\columnwidth]{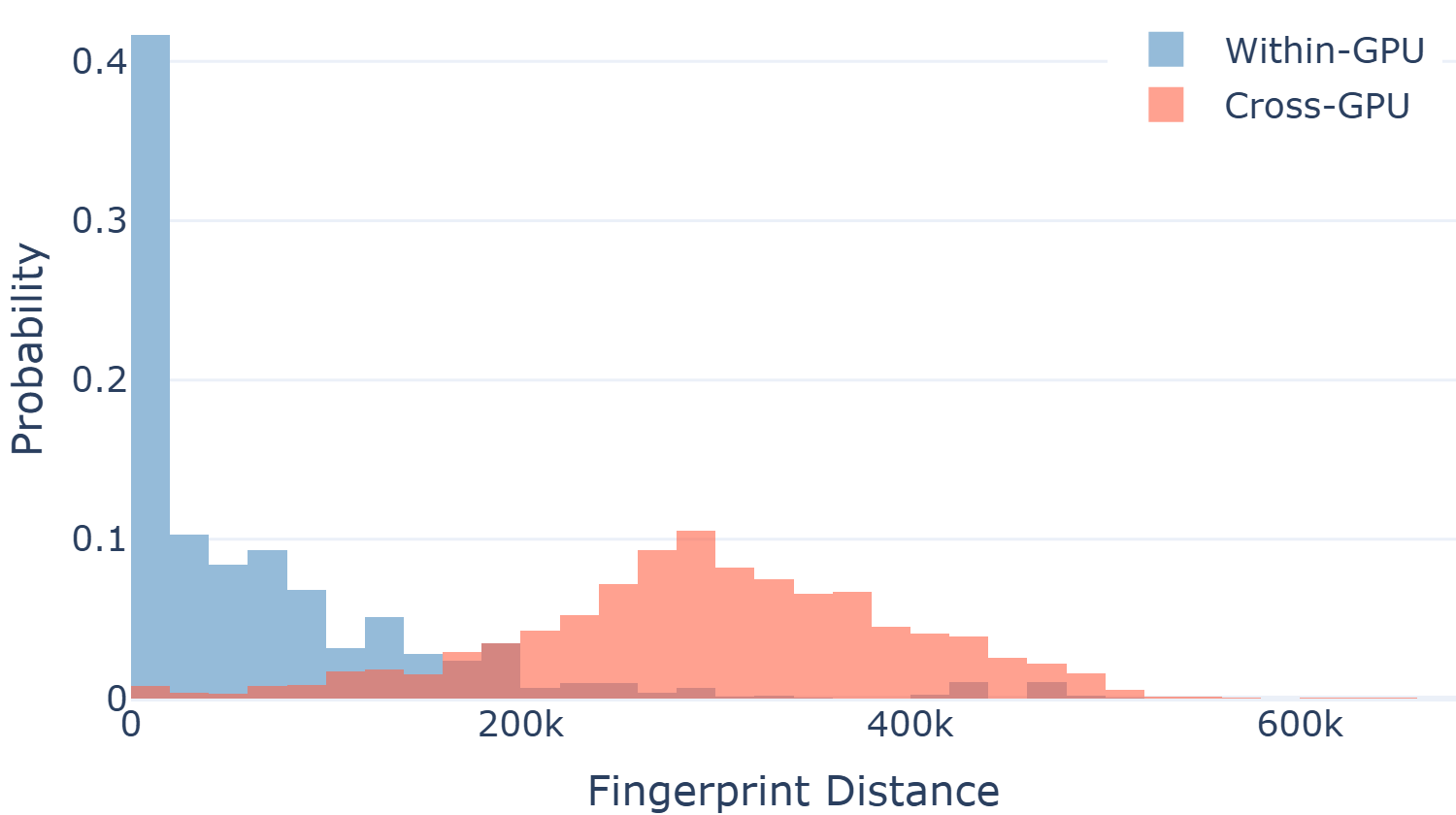}}
    \caption{
      Histogram of within vs. cross-GPU fingerprint distances
    }
    \label{within_vs_cross_gpu}
\end{figure}

We can see that there is good separation between the within vs. cross-GPU distances, showing that fingerprints from the same GPU are generally similar to each other and can be distinguished from those of other GPUs. However, fingerprints from the same GPU do occasionally show large variation. We show ways to overcome this in the next section.

\subsection{Re-Identification Accuracy}
We now consider a more complete setup that includes the registration and verification phases. We do this by splitting the data such that for each GPU and seed, out of 10 runs, 8 are assigned to the registration set, while the remaining 2 are used for verification. Results reported below have been aggregated across possible splits.

We evaluate how accurately we can use the registration set to identify the GPUs of the verification fingerprints. This is known as the re-identification accuracy. We measure this by comparing each verification fingerprint to all registered fingerprints with the same seed to find the closest-matching registered fingerprint. If the associated GPU matches that of the verification fingerprint, then we have successfully identified the GPU. This achieves a re-identification accuracy of 98.8\%.

To further improve accuracy, we can aggregate across multiple runs during the verification phase to reduce noise. For example, we can run the fingerprinting function twice during verification and use only the better matching fingerprint. We simulate this as such: For every pair of fingerprints in the verification set, we compare each fingerprint in the pair to all registered fingerprints. The closest match is the registration fingerprint that has the lowest distance to either of the verification fingerprints. This achieves 100\% re-identification accuracy.

\vskip 0.1in
\begin{table}[h]
  \caption{Summary of results with 95\% confidence intervals
}
  \label{accuracy}
  \begin{center}
    \begin{small}
      \begin{sc}
        \begin{tabular}{lc}
          \toprule
          Approach  & Accuracy  \\
          \midrule
          Single verification run  & 98.8\% (97.3\%, 99.5\%)  \\
          Paired verification runs & 100.0\% (98.5\%, 100.0\%) \\
          \bottomrule
        \end{tabular}
      \end{sc}
    \end{small}
  \end{center}
\end{table}

Our accuracy results are summarized above in \cref{accuracy}. We highlight that our results are achieved without requiring machine learning methods for post-processing or classification of fingerprints. Instead, the fingerprints are compared directly by their distance for improved interpretability.

\subsection{Within vs. Cross-Seed Fingerprint Distances}
Finally, to confirm that the different seeds indeed lead to different fingerprints, we collected the fingerprints of 8 GPUs across 16 seeds, and plot the resulting fingerprint distances within and across seeds, demonstrating clear separation across seeds.

\begin{figure}[h]
    \centerline{\includegraphics[width=\columnwidth]{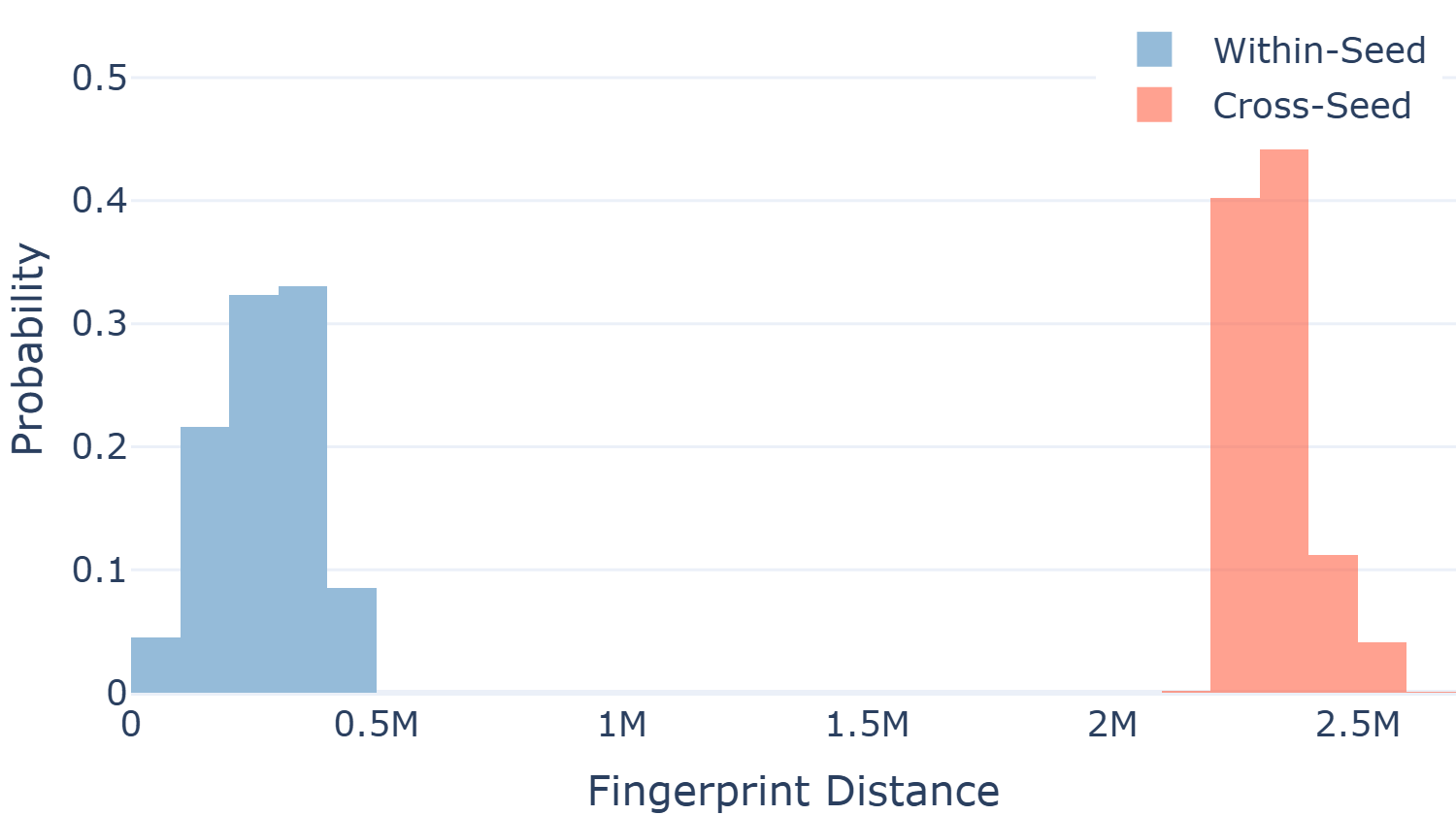}}
    \caption{
      Histogram of within vs. cross-seed fingerprint distances
    }
    \label{within_vs_cross_seed}
\end{figure}

\section{Limitations and Future Work}
We have presented a proof-of-concept fingerprinting function. While the results show good accuracy, more research would be needed before it can be confidently deployed.

\textbf{Scalability.} The function would need to be validated on a larger scale, across more GPUs and seeds to ensure that the fingerprints are sufficiently unique. In order to distinguish among a larger number of GPUs, it is likely that more fingerprints would have to be collected during both the registration and verification phases to maintain the high levels of accuracy.

\textbf{Stability.} It is also important to establish the stability of the fingerprints across time and environmental conditions. This includes ensuring that the fingerprints collected are stable throughout the lifecycle of the GPUs, including after being transported or power cycled, and across a range of temperature conditions. Even if full stability is not achievable, it may be possible to either model the deviations and compensate for them, or record the fingerprints across a range of environmental conditions.

\textbf{Security.} As mentioned above, the protocol is only secure if an adversary cannot optimize the function to run faster, or simulate the GPU well enough and quickly enough to compute the fingerprint without live access to the GPU itself. Establishing this would likely require multiple rounds of iterative red-teaming and further development. This process would be greatly aided by proprietary knowledge of the inner workings of GPUs.

While it may be challenging to fully establish the security of any particular fingerprinting function, we can create an advantage over the adversary by developing multiple different fingerprinting functions. Having a large number of such functions makes it unlikely that all of them would be compromised. Furthermore, even if the adversary has successfully compromised all known fingerprinting functions, they would be deterred by the possibility of more, yet unseen functions that could be sent in the next round of challenges.

\section{Conclusion}
In this paper, we presented a fingerprint-based approach for location verification of advanced GPU chips. This allows for the international monitoring of AI chips without relying on the security of cryptographic keys. We demonstrated a proof-of-concept fingerprinting function that achieved up to 100\% accuracy in small-scale tests. However, further research is needed to establish the security of the algorithm. We hope that continued work on such verification technology will pave the way towards successful international governance of AI development.

\section*{Acknowledgements}
This work was done as part of the Pivotal Research Fellowship. We would like to thank Ariel Gil, as well as the rest of the Pivotal Research team and funders for their instrumental role in enabling this research.

\section*{Impact Statement}
This paper proposes methods to enable more secure and effective AI Governance. While it is important to promote the non-proliferation of dangerous AI capabilities, careful policy design would be needed to ensure that the benefits of AI are accessible to all.

\section*{LLM usage statement}
LLMs were used to generate ideas and experimental code, as well as to edit the writing. All LLM-generated material has been human reviewed.

\bibliography{paper}
\bibliographystyle{icml2026}

\end{document}